\documentstyle[manuscript,aps,epsfig]{revtex}

\begin{document}
\title{The Fourth SM Family Enhancement to the Golden Mode at the Upgraded Tevatron}
\author{O. \c{C}ak\i r$^{a}$ and S. Sultansoy$^{b,c}$}
\address{$^{a}$Ankara University, Faculty of Sciences, Department of Physics, \\
06100, Tandogan, Ankara, Turkey.\\
$^{b}$Gazi University, Faculty of Arts and Sciences, Department of Physics, 
\\
06500, Besevler, Ankara, Turkey.\\
$^{c}$Institute of Physics, Academy of Sciences, H. Cavid Avenue, \\
370143, Baku, Azerbaijan.}
\maketitle

\begin{abstract}
We study the observability for a Higgs boson at upgraded Tevatron via the
modes $gg\rightarrow h\rightarrow ZZ\rightarrow 4l$ ($l=e,\mu $). We find
that the signal can be observed at an integrated luminosity of $30$ fb$^{-1}$
if the fourth SM family exists.
\end{abstract}

In the Standard Model (SM) one doublet of scalar fields is assumed, leading
to the existence of one neutral scalar particle $h$. The requirements of the
stability of \ the electroweak vacuum and the perturbative validity of the
SM allow to set upper and lower bounds depending on the cutoff scale $%
\Lambda $ up to which the SM is assumed to be valid. Experimentally,
constraints on the SM Higgs boson are derived directly from the searches at
LEP2 which lead to $m_{h}>114.3$ GeV \cite{abreu01}. The LHC\ should be able
to cover the full range of theoretical interest up to about $1000$ GeV\cite
{ATLAS99}. A Feynman diagram for the Golden Mode of Higgs production and
decays through heavy quark triangle loop is shown in Fig. \ref{fig1}.

On the other hand, SM does not predict the number of families of fundamental
fermions. In the democratic mass matrix (DMM) approach SM is extended to
include a fourth generation of fundamental fermions with masses typically in
the range from $300$ GeV to $700$ GeV \cite{Datta94},\cite{Celikel95} (for
recent situation see \cite{Salih2000}). The fourth SM family quarks will be
produced copiously at the LHC \cite{ATLAS99}. At the same time a fourth
generation of fermions contributes to the loop-mediated processes in Higgs
production ($gg\rightarrow h$) and decays. In this note we consider the
influence of the fourth SM family on the Higgs boson search at the upgraded
Tevatron.

Two relevant regions of Higgs masses, namely $125-165$ GeV and $175-300$
GeV, require special attention. For Higgs boson mass above $135$ GeV, the
decay mode $h\rightarrow WW$ becomes dominant. Hadronic final state is
owerwhelmed by the QCD background, therefore, one should deal with $W^{\star
}W^{\star }\rightarrow l\nu jj$ and $W^{\star }W^{\star }\rightarrow l\nu
l\nu $ modes \cite{Han99}. However, the channel $h\rightarrow ZZ$ is also
important for the final state observation in the leptonic channel. The decay
width for Higgs boson in the channel $h\rightarrow ZZ$ and its branching
ratio is given in Fig. \ref{fig2}. In the mass range $135-180$ GeV, the
width of the Higgs boson grows rapidly with increasing $m_{h}$.

For Higgs boson masses in the range $175<m_{h}<300$ GeV, the $h\rightarrow
ZZ\rightarrow 4l$ decay mode is the most reliable channel for the discovery
of a SM Higgs boson at the upgraded Tevatron if the fourth SM family exists.
The discovery potential in this channel is primarily determined by the
available integrated luminosity.

The leading production mechanism for a SM Higgs boson at the Tevatron is the
gluon-fusion process via heavy quark triangle loop

\begin{equation}
p\overline{p}\rightarrow ggX\rightarrow hX
\end{equation}

There are also contributions to $h$ production from vector boson fusion
processes, which remain at a low level $(2-10)\%$ comparing to the
gluon-fusion process. Furthermore, gluon fusion process yields the largest
cross section, typically a factor of four above the associated production 
\cite{Han99},\cite{Turcot99}.

The two-loop QCD corrections enhance the gluon fusion cross section by about
80\%. Therefore, we simply rescale the three-level cross sections to match
the NLO result for the overall rate \cite{Spira93}. The results are shown in
Fig. \ref{fig3}. In calculations we have used the CTEQ4M parton distribution
functions \cite{CTEQ}. In the case of three SM families Higgs boson
production cross section is roughly $1.0(0.05)$ pb for $m_{h}=100(300)$ GeV.
However, this cross section is enhanced by the factor $10(6)$ due to the
fourth family quarks. Obviously, the same enhancement takes place for the
Golden Mode and this makes the signal observable over the corresponding
background.

In Fig. \ref{fig4}, we present the cross sections for the process $p%
\overline{p}\rightarrow hX\rightarrow 4lX$ depending on the Higgs mass for
three and four SM family cases. The signal is reconstructed by requiring
four charged leptons $4l$ ($l=e,\mu $) in final state. We use the branching
ratio $B(Z\rightarrow l^{+}l^{-})=3.35\times 10^{-2}$ for $l=e,\mu $. In
Table \ref{table1} we present the number of ``golden'' events in the four SM
family case at integrated luminosity $30$ fb$^{-1}.$

The most serious background is the pair production of Z bosons, $p\overline{p%
}\rightarrow ZZX(Z\rightarrow l^{+}l^{-}),$ which has $\sigma \approx 8$ fb
and should be taken into account for $m_{h}>2m_{Z}$. This background can be
suppressed by consideration of the four-lepton invariant mass distribution.
Invariant mass distributions of four charged lepton final state is given in
Fig. \ref{fig5} for two values of Higgs boson mass. We assume the mass
window of $10$ GeV around the Higgs mass. In the same Figure we present the
main background coming from two Z production. In the four SM family case we
use $m_{4}=640$ GeV. Taking the mass value $m_{4}=320$ GeV leads to
negligible difference in cross sections. As can be seen from the Table \ref
{table2}, we obtain $34$ signal events against $16$ background events within
the mass window $10$ GeV if $m_{h}=200$ GeV. The statistical significance
for $4l$ signal is $8.3$ and $5.4$ for $m_{h}=200$ and $250$ GeV,
respectively. 

In conclusion, the Golden Mode will be observable for $175<m_{h}<300$ GeV
with more than $3\sigma $ significance if the fourth SM family exists. The
same statement takes place also for $125<m_{h}<165$ GeV, however we don't
consider this region in details because it is covered by $h\rightarrow WW$
mode \cite{Han99},\cite{Turcot99}.

\begin{center}
\begin{table}[tbp]
\caption{The expected number of ``golden'' events in the four SM family case
at integrated luminosity $30$ fb$^{-1}.$}
\begin{tabular}{lllll}
\hline
$m_{h}$(GeV) & $\Gamma (h\rightarrow ZZ)$ & B($h\rightarrow ZZ$) & $\sigma
_{4}$(fb) & N($4l$) \\ \hline
$100$ & $3.9\times 10^{-3}$ & $6.8\times 10^{-4}$ & $2.8\times 10^{-2}$ & $%
0.8$ \\ \hline
$110$ & $4.6\times 10^{-3}$ & $2.6\times 10^{-3}$ & $7.8\times 10^{-2}$ & $%
2.\,\allowbreak 3$ \\ \hline
$120$ & $5.6\times 10^{-3}$ & $9.5\times 10^{-3}$ & $2.5\times 10^{-1}$ & $%
7.5$ \\ \hline
$130$ & $7.4\times 10^{-3}$ & $2.5\times 10^{-2}$ & $5.5\times 10^{-1}$ & $%
16.5$ \\ \hline
$140$ & $1.1\times 10^{-2}$ & $4.9\times 10^{-2}$ & $8.9\times 10^{-1}$ & $%
26.7$ \\ \hline
$150$ & $2.0\times 10^{-2}$ & $6.8\times 10^{-2}$ & $1.0$ & $30.0$ \\ \hline
$160$ & $8.3\times 10^{-2}$ & $4.0\times 10^{-2}$ & $4.9\times 10^{-1}$ & $%
14.7$ \\ \hline
$170$ & $3.6\times 10^{-1}$ & $2.3\times 10^{-2}$ & $2.4\times 10^{-1}$ & $%
7.\,\allowbreak 2$ \\ \hline
$180$ & $6.1\times 10^{-1}$ & $5.9\times 10^{-2}$ & $5.0\times
10^{-1}\allowbreak $ & $15.0$ \\ \hline
$190$ & $9.9\times 10^{-1}$ & $2.1\times 10^{-1}$ & $1.4$ & $42.0$ \\ \hline
$200$ & $1.4$ & $2.5\times 10^{-1}$ & $1.5\allowbreak $ & $45.0$ \\ \hline
$220$ & $2.3$ & $2.8\times 10^{-1}$ & $1.1\allowbreak $ & $33.0$ \\ \hline
$240$ & $3.4$ & $2.9\times 10^{-1}$ & $8.4\times 10^{-1}\allowbreak $ & $25.2
$ \\ \hline
$260$ & $4.7$ & $3.0\times 10^{-1}$ & $6.3\times 10^{-1}\allowbreak $ & $18.9
$ \\ \hline
$280$ & $6.4$ & $3.0\times 10^{-1}$ & $4.7\times 10^{-1}\allowbreak $ & $%
14.\allowbreak 1$ \\ \hline
$300$ & $8.4$ & $3.1\times 10^{-1}$ & $3.6\times 10^{-1}$ & $\allowbreak
10.\,8$ \\ \hline
\end{tabular}
\label{table1}
\end{table}

\begin{table}[tbp]
\caption{Observability of the ``golden'' events for integrated luminosity $30
$ fb$^{-1}$ within the mass window $10$ GeV. \ }
\begin{tabular}{cccccc}
\hline
& \multicolumn{2}{c}{S} & B & \multicolumn{2}{c}{S/$\sqrt{\text{B}}$} \\ 
\hline
$m_{h}$(GeV) & SM-3 & SM-4 &  & SM-3 & SM-4 \\ \hline
$200$ & $3.5$ & $33.8$ & $16.4$ & $0.9$ & $8.3$ \\ \hline
$250$ & $2.0$ & $16.0$ & $8.8$ & $0.7$ & $5.4$ \\ \hline
\end{tabular}
\label{table2}
\end{table}

\begin{figure}[tbp]
\epsfig{file=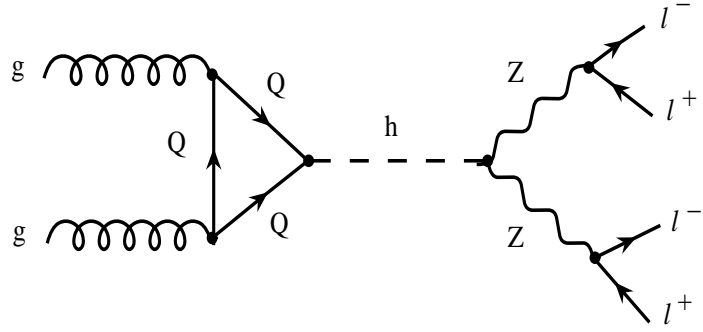,width=10cm,height=5cm}
\caption{Feynman diagram for the Golden Mode $gg\rightarrow h\rightarrow
ZZ\rightarrow 4l$ at Tevatron.}
\label{fig1}
\end{figure}

\begin{figure}[tbp]
\epsfig{file=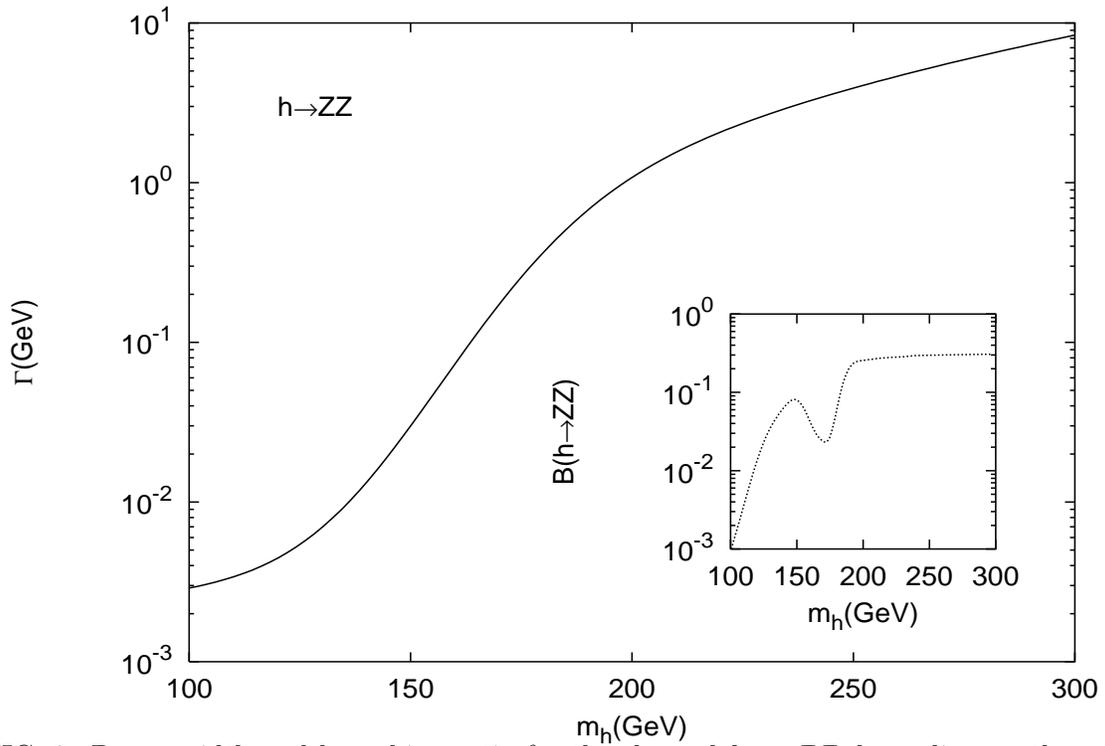,width=15cm,height=10cm}
\caption{Decay width and branching ratio for the channel $h\rightarrow ZZ$
depending on the mass of Higgs boson. }
\label{fig2}
\end{figure}

\begin{figure}[tbp]
\epsfig{file=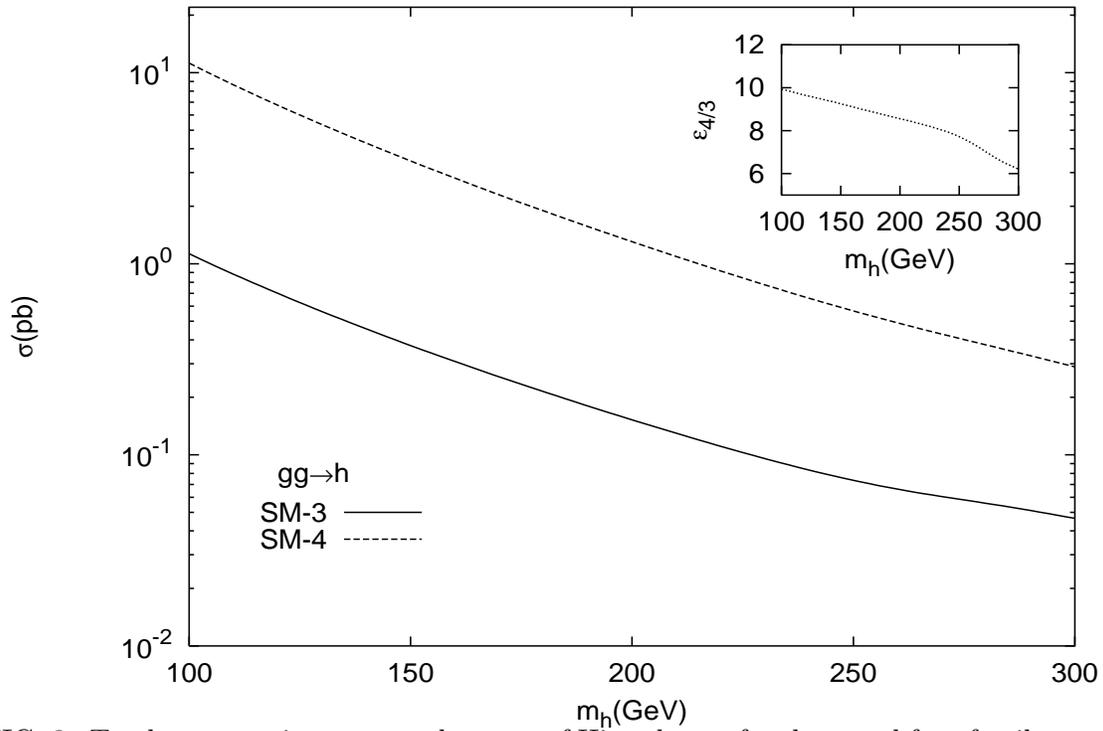,width=15cm,height=10cm}
\caption{Total cross sections versus the mass of Higgs boson for three and
four family cases. An enhancement factor $\protect\varepsilon _{4/3}$
depending on the Higgs mass is also shown. }
\label{fig3}
\end{figure}

\begin{figure}[tbp]
\epsfig{file=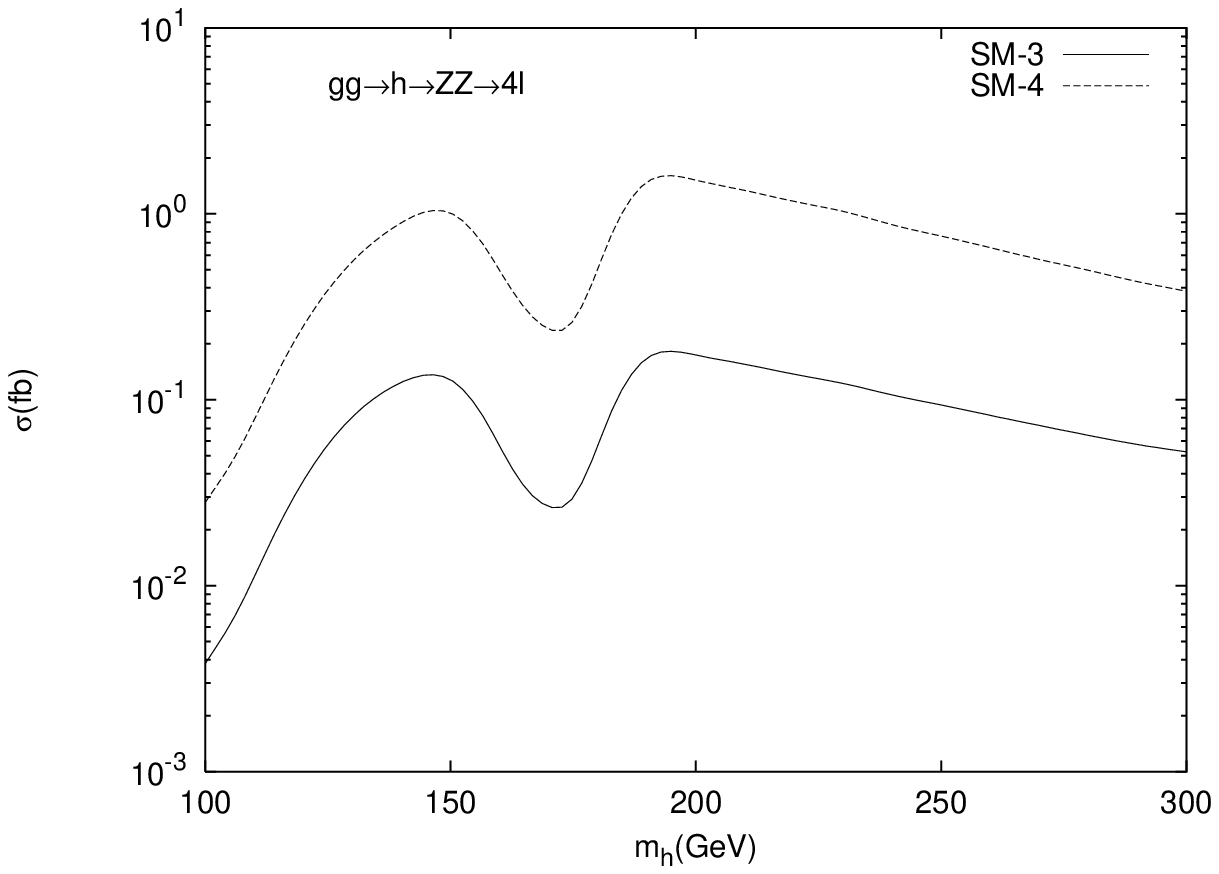,width=15cm,height=10cm}
\caption{The cross section of ``golden'' events depending on the Higgs mass
for the cases of three and four SM families.}
\label{fig4}
\end{figure}

\begin{figure}[tbp]
\epsfig{file=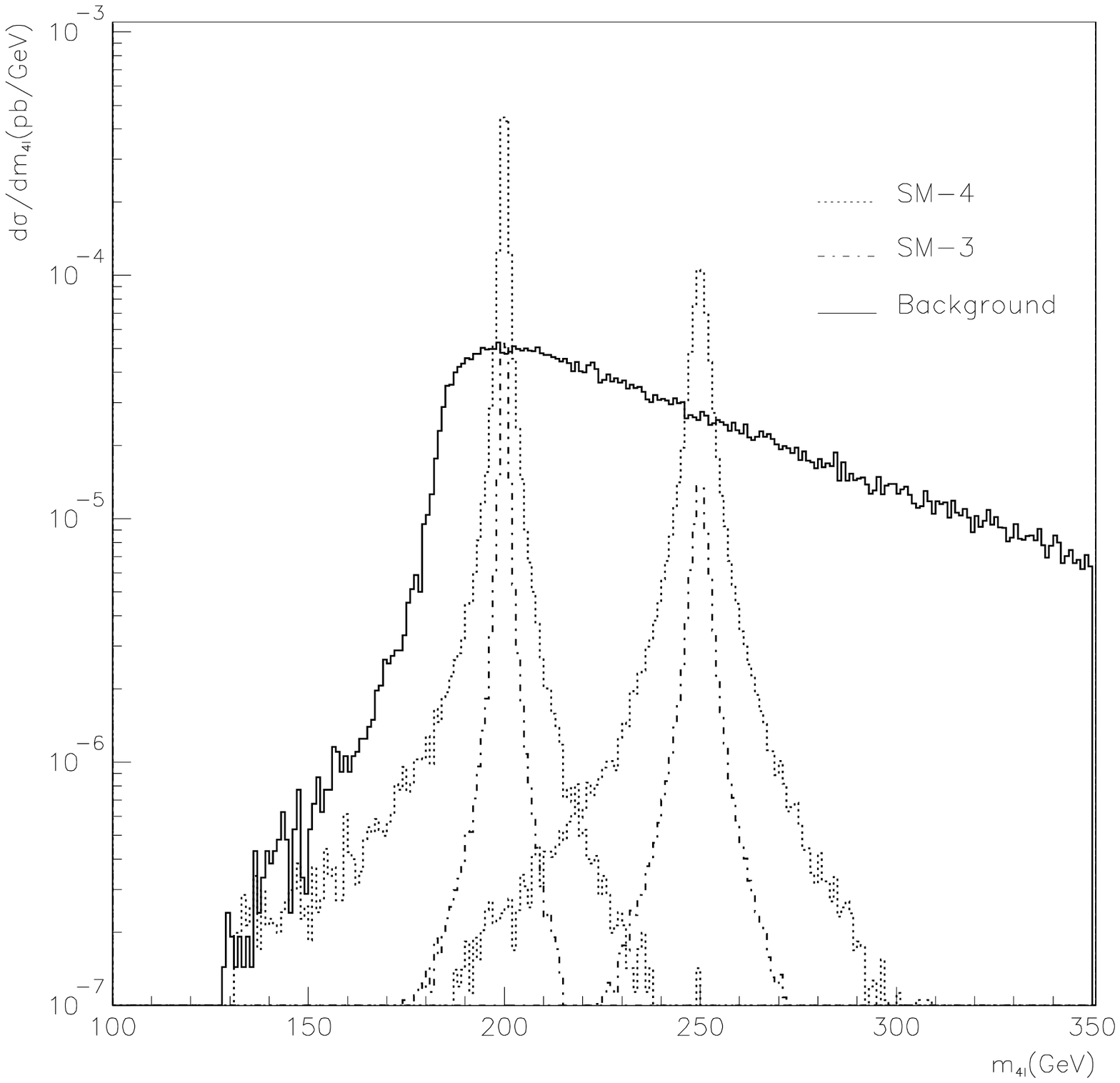,width=15cm,height=10cm}
\caption{The invariant mass distributions of the signal and background for $%
m_{h}=200$ and $250$ GeV.}
\label{fig5}
\end{figure}
\end{center}


\begin{references}
\bibitem{abreu01}  P. Abreu {\it et al.,} DELPHI Collaboration, Phys. Lett.
B499 (2001) 23.

\bibitem{ATLAS99}  ATLAS Collaboration, ATLAS Technical Design Report,
CERN/LHCC-99-15, (1999).

\bibitem{Datta94}  A. Datta and S. Raychaudhuri, Phys. Rev. D49 (1994) 4762.

\bibitem{Celikel95}  A. \c{C}elikel, A.K. \c{C}ift\c{c}i and S. Sultansoy,
Phys. Lett. B342 (1995) 257.

\bibitem{Salih2000}  S. Sultansoy, hep-ph/0004271.

\bibitem{Han99}  Tao Han and Ren-Jie Zhang, Phys. Rev. Lett., 82, (1999)
25-28.

\bibitem{Turcot99}  Tao Han, A. S. Turcot and Ren-Jie Zhang, Phys. Rev. D59,
(1999) 093001.

\bibitem{Spira93}  D. Graudenz, M. Spira and P.M. Zerwas, Phys. Rev. Lett.
70, (1993) 1372; M. Spira, A. Djouadi, D. Graudenz and P.M. Zerwas, Nucl.
Phys. B453, (1995) 17.

\bibitem{CTEQ}  CTEQ Collaboration, H.L. Lai {\it et al., }Phys.Rev. D55,%
{\it \ (1997) 1280.}
\end{references}
\end{document}